\documentclass[a4paper,12pt]{article}
\usepackage{graphicx,amssymb,color}
\usepackage{epsfig}
\usepackage{color}

\begin{document}

\centerline{\LARGE \bf Modelling of cell choice between differentiation}

\medskip

\centerline{\LARGE \bf
and apoptosis on the basis of intracellular and}

\medskip

\centerline{\LARGE \bf
extracellular regulations and stochasticity}

\vspace*{1cm}

\centerline{\bf M. Banerjee$^1$, V. Volpert$^2$}

\vspace*{0.5cm}

\centerline{$^1$ Department of Mathematics and Statistics,
Indian Institute of Technology Kanpur}

\centerline{Kanpur 208016, India, E-mail: malayb@iitk.ac.in}

\centerline{$^2$ Institut Camille Jordan, UMR 5208 CNRS, University Lyon 1}

\centerline{69622 Villeurbanne, France, E-mail:
volpert@math.univ-lyon1.fr}

\vspace*{1cm}

\noindent {\bf Abstract.} The work is devoted to the analysis of
cell population dynamics where cells make a choice between
differentiation and apoptosis. This choice is based on the values of
intracellular proteins whose concentrations are described by a
system of ordinary differential equations with bistable dynamics.
Intracellular regulation and cell fate are controlled by the
extracellular regulation through the number of differentiated cells.
Initial intracellular protein concentrations are considered for each
cell as random variables with a given area of variation.
Intracellular regulation, extracellular regulation and random
initial conditions work together to produce differentiated cells and
to control their number. The role of intracellular regulation is to
provide a possible choice between differentiation and apoptosis,
extracellular regulation controls the number of differentiated
cells, stochastic initial conditions can suppress oscillations and
provide stability of the system.\\

\noindent {\bf Key words:} Cell differentiation, apoptosis, extra
cellular regulation, cell fate.

\vspace*{1cm}

%%%%%%%%%%%%%%%%%%%%%%%%%%%%%%%%%%%%%%%%%%%%%%%%%%%%%%%%

\section{Introduction}

\subsection{Cell fate and multi-scale modelling}

Cell population dynamics is determined by the equilibrium between self-renewal, differentiation and apoptosis.
Cell fate depends on complex intracellular and extracellular regulations which can act on the level of the
whole cell population and not only on individual cells \cite{lei}, \cite{macarthur}, \cite{macarthur2}, \cite{morris}.
Multi-scale modelling provides an appropriate tool to study cell population dynamics.
There are many different interpretations of multi-scale modelling in biology, as well as many models and applications
(see \cite{Anderson2007}, \cite{Bernard}, \cite{Cristini}, \cite{Osborne}, \cite{v2014} and the references therein).
In the case of physiological processes, by multi-scale models we will understand  the models which consider some populations of cells (tissue),
intracellular regulation of cell fate and possibly of some its properties, extracellular regulation by the surrounding cells and/or by
other organs and tissues.

We will distinguish local and
global extracellular regulation. By local regulation we will understand regulation from the surrounding cells of the same tissue.
It can occur by the direct cell-cell contact or by intermediate of various molecules produced by cells of the considered tissue and diffusing in the extracellular matrix. In particular local regulation occurs in stem cell niche. Stem cells can self-renew only being in contact with other cells in the niche
(e.g., stromal cells). When they lose this contact, they differentiate and cannot self-renew any more. This local control of self-renewal is important to prevent uncontrolled self-renewal leading to cancer.

Global extracellular regulation is effectuated by other organs and tissues. In particular, this can be hormones, nutrients, growth factors.
Global regulation takes into account some overall information about the tissue and ignores the details about local cell distribution inside it.
This information can be based on the number of cells of certain type. For example, production of the hormone erythropoietin in the kidney
depends on the number of erythrocytes in blood (more precisely, on hemoglobin). Erythropoietin arrives to the bone marrow and regulates production
of erythrocytes \cite{kb1990}. We will return to this example below.

%%%%%%%%%%%%%%%%%%%
%
%\subsection{Hybrid models and hematopoiesis}

Hybrid discrete-continuous models are well adapted for multi-scale modelling in biology. They consider cells as individual objects, intracellular regulatory networks can be described by ordinary differential equation and biochemical substances in the extracellular matrix by partial differential equations.
They can be completed by various models of global extracellular regulation. Such models are applied to study various physiological processes \cite{Glade}
and dynamics of cell populations \cite{Bessonov}, \cite{Bessonov2012}, \cite{v2013}.

Among other applications, hybrid models are developed to study hematopoiesis (blood cell production) \cite{Bessonov}, \cite{bcfkv}.
Hematopoiesis represents an interesting example to study dynamics of cell populations. It is a complex process which begins
with hematopoietic stem cells and results in production of red blood cells, platelets and leucocytes. During this process, cells undergo many
consecutive steps of transformation controlled by numerous local and global regulations. The model developed and studied
in this work is inspired by hematopoiesis but it admits also some other interpretations.

Traditionally mathematical models of hematopoesis are based on ordinary and delay differential equations and on structured cell dynamics
(transport equations). Dynamics of hematopoietic stem cells was studied in \cite{m1978}, \cite{m1979} (see also \cite{roeder2006} and references therein).
Periodic oscillations in blood cell count and various related blood diseases were studied in \cite{am2008}, \cite{bbm2003}, \cite{{cm2005b}}, \cite{santillan-etal2000}. A model of platelet production was developed in \cite{eller-etal1987}, \cite{wichmann-etal1979}.
Mathematical modelling of erythropoiesis and anemia began in \cite{wl1985}, \cite{wlpw1989},  \cite{wwlp1989}.
Age-structured models of erythropoiesis which take into account the action of erythropoietin are considered in \cite{adit2006}, \cite{crauste2008}, \cite{mbm1998}. Interaction of different cell lineages was studied in \cite{cm2005a},  \cite{cm2005b}. There are numerous works devoted to modelling of leukemia development and treatment \cite{Bessonov2009}, \cite{halanay1}, \cite{halanay2}, \cite{ozbay}, \cite{stiehl}.

It is important to indicate that in all these works the rates of self-renewal,  differentiation and apoptosis are imposed as given constants or functions
of some concentrations. However usually these values are not known or, what is more important, they even cannot be considered as given.
Indeed, for the same concentrations of hormones, drugs, nutrients and other extracellular substances their action can depend on
the concentrations of intracellular  proteins and, more general, on the cell state and history. Therefore we need to take
into account intracellular regulation and to introduce multi-scale models.

Multi-scale hybrid models of hematopoiesis were developed in  \cite{AML2012}, \cite{crauste2008}, \cite{demin}, \cite{Fischer},
\cite{jtb},  \cite{SIAM2011}, \cite{acta}. These models descrbe self-renewal, differentiation
and apoptosis of erythroid progenitors on the basis of intracellular and extracellular regulation. The intracellular regulation
takes into account the main proteins involved in this process. The extracellular regulation is based on the
interaction of Fas-lignad producing cells which stimulate differentiation and apoptosis, on the interaction with macrophages
which promote self-renewal of erythroid progenitors and on the influence of erythropoietin which downregulates apoptosis of
erythroid progenitors. This approach allowed a detailed description of erythropoiesis in normal and pathological situations.

\subsection{About this work}

\paragraph{Biological motivation.}

Proliferation and differentiation of hematopoietic stem cells give all lineages of blood cells, erythrocytes, platelets and
several lineages of white blood cells. Each of them begins with progenitors, immature cells which can differentiate, die by apoptosis and some
of them self-renew. The equilibrium between these three possibilities has a crucial importance for normal functioning of hematopoiesis.
If it is not preserved, various blood diseases including leukemia can develop.

Let us consider this process in more detail with the example of erythropoiesis, red blood cell production. This lineage begins with erythroid progenitors
(colony forming units) which can self-renew, differentiate or die by apoptosis. They form erythroblastic islands, small units with several dozens
of cells around a macrophage \cite{eymard}, \cite{Fischer}. Their self-renewal occurs only when they are close to macrophage, otherwise they differentiate or die. This confirms the observation that self-renewal can take place only under a tight local regulation. Only global regulation is not sufficient. In this
case erythroblastic islands can have unbounded growth or extinction \cite{AML2012}. Differentiation and apoptosis of erythroid progenitors is determined by the concentration of Fas-ligand \cite{maria}. It is a growth factor produced either by the progenitors themselves (mice) \cite{socol} or by more mature cells (humans) \cite{chasis2008}.

Successive differentiation of erythroid progenitors gives proerythroblast, erythroblasts and finally reticulocytes which leave the bone marrow
into blood flow where they become mature erythrocytes. In normal case, these cells can only differentiate or die by apoptosis, but not self-renew.
The latter can occur due to genetic mutations resulting in erythroleukemia \cite{Bessonov2009}. The choice between differentiation and apoptosis of these cells depends on the level of the hormone erythropoietin produced in the kidney. The rate of its production depends on the relation between the quantity
of oxygen required by the organism and the global quantity of hemoglobin in blood. In the first approximation we can relate the quantity of hemoglobin to the number of erythrocytes. Erythrocytes have a given life span in blood, after which they die. It is about 40 days for mice and 120 days humans.

Thus, simplifying this description, we can summarize it as follows. Erythroblasts can either differentiate and give mature erythrocytes or die
by apoptosis. The rates of differentiation and apoptosis depend on the number of erythrocytes in blood. We do not consider here erythroid progenitors
because they can also self-renew, and we omit intermediate staged of erythroblast maturation. Similar processes occur in other cell lineages
of hematopoiesis and in other tissues. We will use this schematic description as biological motivation for the model considered in this work.

\paragraph{Model and objectives.}

The purpose of this work is to suggest a relatively simple model of cell population dynamics which would allow us to study
how intracellular and extracellular regulations and stochasticity influence cell fate. In order to simplify the model, we will deliberately not
take into account here spatial heterogeneity and local cell-cell interactions. As discussed above, this is appropriate for some stages of
hematopoiesis and for other tissues.

Hence we will consider only two cell types. Cells of the first type can differentiate or die by apoptosis. When they differentiate, they
produce cells of the second type whose number determines the rates of differentiation and apoptosis of cells of the first type. Such models
are often considered in modelling of hematopoiesis (Section 1.1) with the only difference that differentiation and apoptosis will be
determined here by the intracellular regulation and will not be considered as given constants or functions. This difference is crucial.
%This does not mean that the models where cell fates are given are wrong. We need to begin with a more complete model and to find the
%conditions when it can be reduced to the simpler model.
Hybrid models of hematopoiesis presented above take into account
intracellular regulation. However they are developed to study functioning of erythroblastic islands with self-renewal of erythroid progenitors,
local cell interactions, cell motion and other factors. We need a simpler model to study cell fate in a more pure and general setting.
We will present it in more detail in the next section.

Thus, we will consider in this work a hybrid multi-scale model with intracellular regulation and global extracellular regulation.
We will describe the intracellular regulation with systems of ordinary differential equations for the concentrations of intracellular proteins. Therefore we need to specify the initial values of these concentrations for each cell. If these initial concentrations are the same for all cells then,
under constant external conditions (which is the case for a steady state solution), the fate of all cells will also be the same.
Hence in order to describe cell choice between differentiation and apoptosis, we need to introduce some random variations in the initial conditions. This is one of the important features of the model. We will see below that in some cases randomness of initial conditions influences the behavior of the system. It can suppress oscillations and stabilize the system.

Let us note that random perturbations can occur not only for the initial conditions but also during the whole cell life.
In this case, we need to consider stochastic differential equations for the intracellular regulation (see \cite{Golubev} and the references therein). We will restrict ourselves here to the case of random initial conditions for two reasons. First of all, to simplify the model and to separate the effects of random initial conditions from random perturbations during whole cell evolution. Second, in bistable dynamics, when the trajectory is already inside the basin of
attraction of one of stable stationary points, small random perturbations are not so important. They are more essential in the beginning
of this evolution when the trajectory is close to the manifold separating basins of attraction. In this case, it is similar to the introduction of random initial conditions.

Another key feature of the model will be the presence of a global regulation of the number of differentiated cells. If the number of cells in the population
at time $t$ is $N(t)$ and the target number which should be achieved by the system is $N_0$, then the system will adapt its parameters depending
on the difference $N(t)-N_0$. We will show that under some conditions behavior of the system is determined by the global regulation and its
dependance on the intracellular regulation and on stochasticity is weak. In this case we can model cellular systems with an uncomplete
information about intracellular regulation and about mechanisms and parameters of possible stochasticity.

Thus, the main goal of this work is to show how intracellular regulation, global extracellular regulation and random initial conditions
determine evolution of cell populations. We can summarize their action as follows. Intracellular regulation provides a choice of cell fate, extracellular regulation controls this choice, random initial conditions stabilize the system.
We will introduce the model in the next section. Numerical simulations will be presented in Section 3 and an approximate analytical model
in Section 4.

%%%%%%%%%%%%%%%%%%%%%%%%%%%%%%%%%%%%%%%%%%%%

%\newpage

\section{Model}

\begin{figure}[htbp]
\centerline{\includegraphics[scale=0.6]{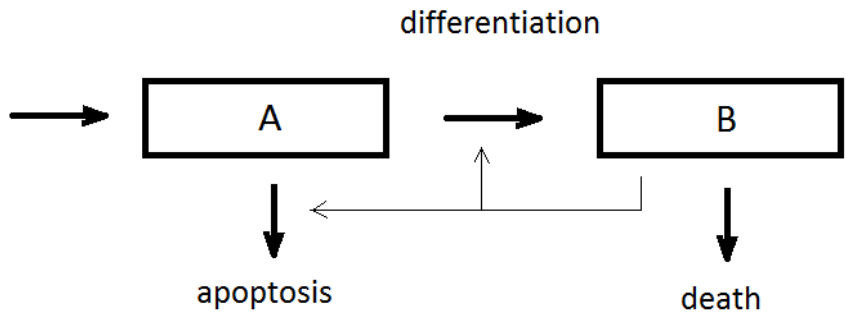}}
\caption{Schematic representation of the model. Cells of the type $A$ can differentiate into cells $B$
or die by apoptosis. Cells $B$ have a fixed life span after which they die. The number of differentiated cells $B$
influence the choice of cells $A$ between differentiation and apoptosis. This choice is determined by intracellular
proteins described by a system of ordinary differential equations.}
\label{scheme}
\end{figure}

We will consider cells of two types, $A$ and $B$.
Every given time interval $\tau$ a new cell $A$ appears due to differentiation of less mature cells
into cells $A$. Cells of the type $A$ can either differentiate into cells
of the type $B$ or, otherwise, they die by apoptosis. Hence we restrict ourselves
here to the cell choice between differentiation and apoptosis. We do not consider here self-renewal of cells $A$ in order
 to study first the functioning of the system in this simpler case. Self-renewal takes place for more immature cells
 which provide a constant influx of cells $A$. This model is shown schematically in Figure \ref{scheme}.

Cells of the type $B$ have a finite and fixed life span $T$. After that they die. The total number $N(t)$ of cells $B$
at time $t$ is determined by the number of cells $A$ differentiated during the time interval $[t-T,t]$.
Suppose that the system should possess $N_0$ cells $B$ for the normal functioning of the organism. Then
 the rate of differentiation of cells $A$ should depend on $N(t)$ and $N_0$.

%We will present the model in more detail below in this section. Let us note that such regulation is specific in particular
%for hematopoiesis. There are numerous works on modelling of normal and pathological hematopoiesis where similar models were
%considered with ordinary or delay differential equations [].

%Let us begin with the simplest model which describes dynamics of this cell population.
%If we introduce concentrations of cells $A$ and $B$ and use for them the same notation,
%then we can write the system of two ordinary differential equations:
%
%\begin{equation}\label{sys1}
%  \frac{dA}{dt} = a - (k_d + k_a) A , \;\;\;\;
%  \frac{dB}{dt} =  k_d A -  \sigma B ,
%\end{equation}
%where $a$ is the influx of cells $A$, $k_a$ and $k_d$ are the rates of their differentiation and apoptosis,
%$\sigma$ is the mortality coefficient of cells $B$. Assuming that the coefficients $k_d$ and $k_a$ depend on $B$,
%$k_d(B)$ is a decreasing function and $k_a(B)$ increases, then this system of equations has a unique stationary point
%$(A_0,B_0)$, where
%
%$$ B_0 = \frac{a}{\sigma} \; \frac{k_d(B_0)}{k_a(B_0)+k_d(B_0)} \; , \;\;\;\;
%A_0 = \frac{\sigma B_0}{k_d(B_0)} \; . $$
%This stationary point is globally asymptotically stable. Let us note that differentiation of cells $A$ is not instantaneous.
%If we take this into account, we obtain delay differential equations.

In general, the rates of differentiation and apoptosis cannot be considered as given constants or functions.
Indeed, cell response on the extracellular regulation can depend on its intracellular regulation. Therefore we need
to introduce this regulation in the model. We will study the model with intracellular regulation in Section 3.
When we have this more complete model, it can be possible to formulate some simplifying assumptions
under which it can be reduced to delay differential equations (Section 4).

%In this work we will study the case where the cell fate is determined by concentrations of intracellular proteins $p_1, p_2$ described
%by ordinary differential equations. Instead of system (\ref{sys1}) we need to consider either transport equations for the
%concentrations  $A(p_1,p_2,t)$, $B(p_1,p_2,t)$ or individual cells. We will use the second approach.

\paragraph{Intracellular regulation.}

%We consider a system of ordinary differential equations for intracellular proteins whose concentrations determine cell fate.
We consider here the simplest regulation with only two proteins $p_1$ and $p_2$ and, respectively, two equations
for their concentrations:

\begin{equation}\label{mod1}
  \frac{dp_1}{dt} = F_1(p_1,p_2) , \;\;\;\;
  \frac{dp_2}{dt} = F_2(p_1,p_2) \; .
\end{equation}
Each cell  has its own values $p_1(t)$ and $p_2(t)$ described by this system of equations. Obviously, these
concentrations can be different in different cells.

When a new cell $A$ appears, we prescribe it some initial values $p_1^0$ and $p_2^0$. Hence we can determine the evolution of these
concentrations inside each cell.
We suppose that if the concentration $p_1(t)$ reaches some critical level $p_1^*$, then the cell dies by apoptosis. If
the concentration $p_2(t)$ reaches some critical level $p_2^*$, then the cell differentiates.

\begin{figure}[htbp]
\centerline{%\includegraphics[scale=0.6]{fig-intra.ps}
\includegraphics[scale=0.45]{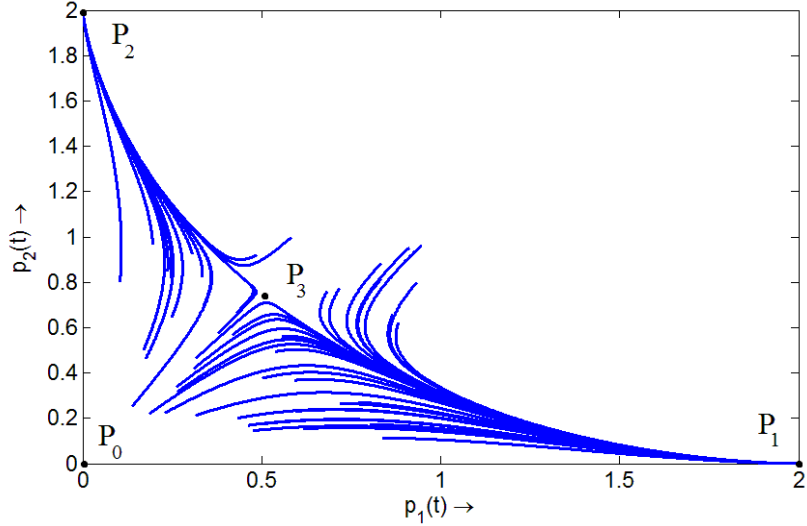}}
\caption{Trajectories of system (\ref{mod1}). }
\label{intra}
\end{figure}

Evolution of the intracellular concentrations depends on the functions $F_1$ and $F_2$. We will consider them in the following form:

\begin{equation}\label{mod2}
  F_1(p_1,p_2) = k_1 p_1 (1 -a_{11} p_1 - a_{12} p_2) , \;\;\;\;
   F_2(p_1,p_2) = k_2 p_2 (1 -a_{21} p_1 - a_{22} p_2) ,
\end{equation}
where $k_1$, $k_2$ and $a_{ij}$ are positive parameters.
%Such systems were introduced in population dynamics by Lotka \cite{Lotka} and Volterra \cite{Volterra} and then used by many authors.
We consider the functions $F_1$ and $F_2$ in this form in order to describe bistable dynamics. They can be different
depending on applications but their specific form is not essential for what follows.
This system can have up to four
stationary points with non-negative coordinates: $P_0=(0,0)$, $P_1=(1/a_{11},0)$,
$P_2=(0,1/a_{22})$, $P_3 = (\tilde p_1, \tilde p_2)>0$, where $\tilde p_1$ and  $\tilde p_2$ satisfy the system
of equations

$$ a_{11} p_1 + a_{12} p_2 = 1 , \;\;\;\; a_{21} p_1 + a_{22} p_2 = 1 . $$
We suppose that it has a positive solution, and the points
$P_1$ and $P_2$ are stable, while the point $P_3$ is unstable.
These conditions are satisfied if $a_{21} > a_{11} > 0$, $a_{12} > a_{22} > 0$.
The separatrix of the point $P_3$
separates the basins of attraction of the points $P_1$ and $P_2$. If the initial condition is below this line,
then the trajectory converges to the point $P_1$, if the initial
condition is above this line, then the trajectory converges to the
point $P_2$. If the trajectory approaches the point $P_1$, then the concentration $p_1$ becomes
large, the concentration $p_2$ small. In this case the cell will
die by apoptosis. In the second case, $p_2$ becomes large, $p_1$ small, the cell differentiates.
Figure \ref{intra} shows behavior of trajectories of this system for random initial conditions chosen in $[0.1,1] \times [0.1,1]$.

\paragraph{Cell number and extracellular regulation.}

The number of cells $B$ at time $t$ is denoted by $N(t)$. It equals the number of cells $A$ differentiated
during the time interval from $t-T$ to $t$. Suppose that the system should produce $N_0$ cells. If $N(t)$ is different
from $N_0$, then there is a feedback control which acts on the rate of differentiation. Since the choice between differentiation
and apoptosis of cells $A$ is determined by the concentrations $p_1$ and $p_2$ described by system (\ref{mod1}),
we will suppose that its coefficients depend on the difference $N(t)-N_0$. Namely, we set

\begin{equation}\label{mod3}
  a_{21}(N) = a_{21}^0 + \alpha(N-N_0) .
\end{equation}
In order to reduce the number of parameters, we consider a linear dependence of the coefficient on $N$.
We verify during the simulations that it remains positive.
All other coefficients are independent of $N$.

If $N(t) > N_0$, then the coefficient $a_{21}$ increases in
comparison with $a_{21}^0$. Therefore the function $F_2(p_1,p_2)$ in
(\ref{mod2}) decreases. Hence growth of the concentration $p_2$
decelerates and, consequently, the rate of differentiation also
decreases. Thus the dependence of the coefficients of the
intracellular regulation on the cell number provides a global
control over the system with the purpose to get the required number
of differentiated cells.

\paragraph{Stochasticity in the initial conditions.}

Let us recall that there is a constant influx of cells $A$. This means that every time interval $\tau$ a new cell enters the compartment
with cells $A$. We need to prescribe the initial concentrations $p_1^0$ and $p_2^0$ for each new cell. Since these values are usually
not known and they can hardly be measured experimentally, this becomes an additional unknown parameter which can influence
behavior of the system. We will choose the initial conditions from the square domains

\begin{equation}\label{mod4}
  D = \{ a - k \leq p_1^0 \leq a + k , \;\;\; b - k \leq p_2^0 \leq b + k \} ,
\end{equation}
where $a$, $b$ and $k$ are chosen in such a way that the whole domain $D$ has non-negative coordinates.
The initial condition will be chosen from the domain $D$ as a random variable with a uniform distribution.

%%%%%%%%%%%%%%%%%%%%%%%%%%%%%%%%

%\newpage

\setcounter{equation}{0}

\section{Numerical results}

We present results of numerical simulations of the model described in Section 2. We set
$\tau=0.1$, $T=20$, $a_{11}=a_{22}=0.5$, $a_{12}=a_{21}^0=1$, $p_1^*=p_2^*=1.8$ and we will vary other parameters.

\subsection{Deterministic initial conditions}

We begin with the case where the initial conditions for the concentrations $p_1^0$, $p_2^0$ are fixed.
Therefore all cells $A$ have exactly the same initial conditions in the intracellular regulation.
Then this problem does not have solutions with a constant value $N(t) \equiv N$. In this sense we can say that the
problem with deterministic initial conditions does not have stationary solution.
Indeed, suppose that such solution exists and consider the phase plane of the system (\ref{mod1}).
Since $a_{21}(N)$ is constant (time independent) then the separatrix which separates basins of attraction of the points
$P_1$ and $P_2$ is also fixed. If the initial condition $(p_1^0,p_2^0)$ belongs to the basin of attraction of the point
$P_1$, then all cells will die. If it belongs to the basin of attraction of the point $P_2$, then all cells will differentiate.
Finally, if the initial condition is exactly at the sepratrix, the corresponding trajectory will converge to the unstable
point $P_3$ and it will never reach the critical values $p_1^*$ or $p_2^*$. Hence in all three cases the number of differentiated
cells cannot be equal $N$ if it is different from $0$ or $T/\tau$.

\begin{figure}[htbp]
\centerline{\includegraphics[scale=0.37]{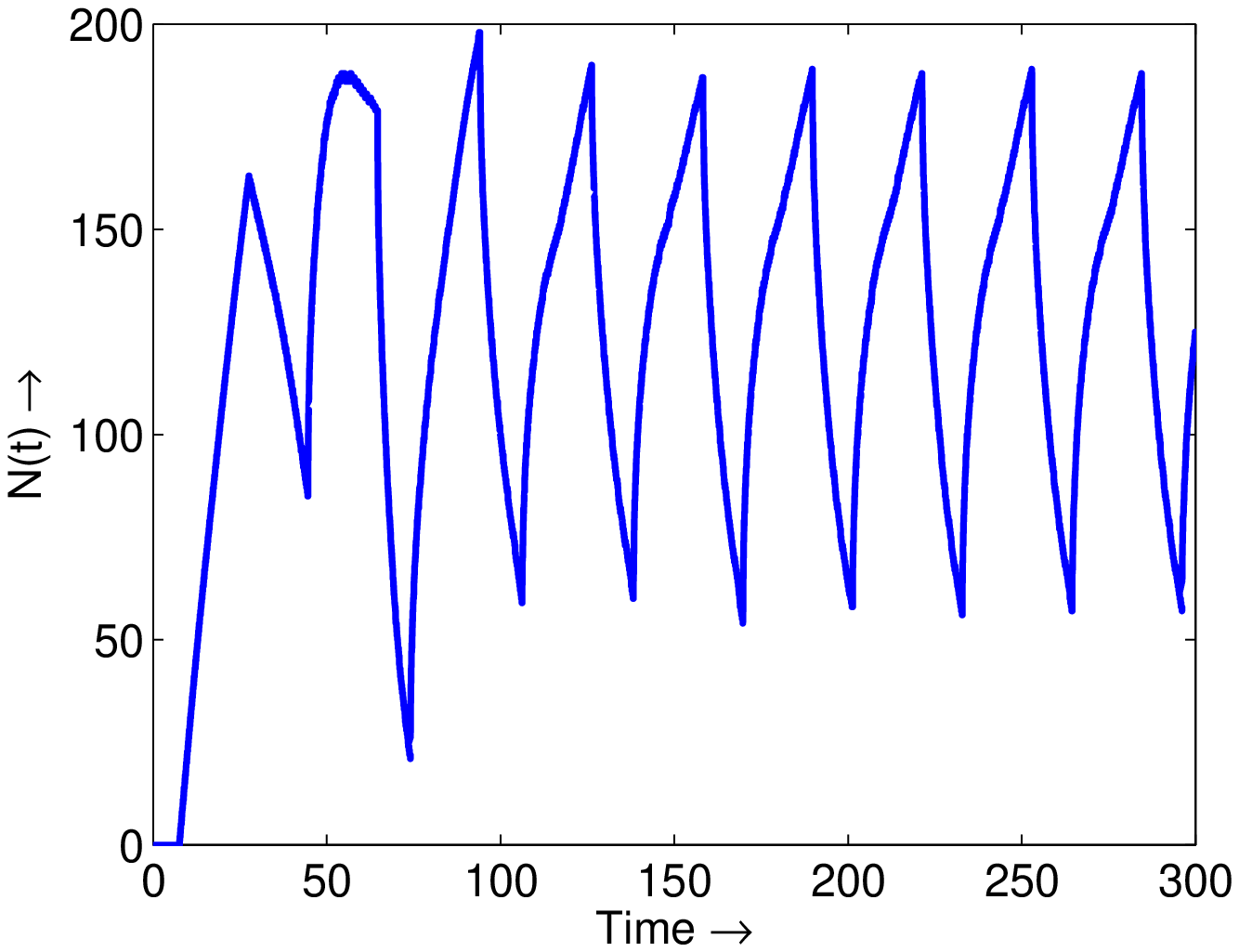}
 \hspace*{-0.5cm}
\includegraphics[scale=0.37]{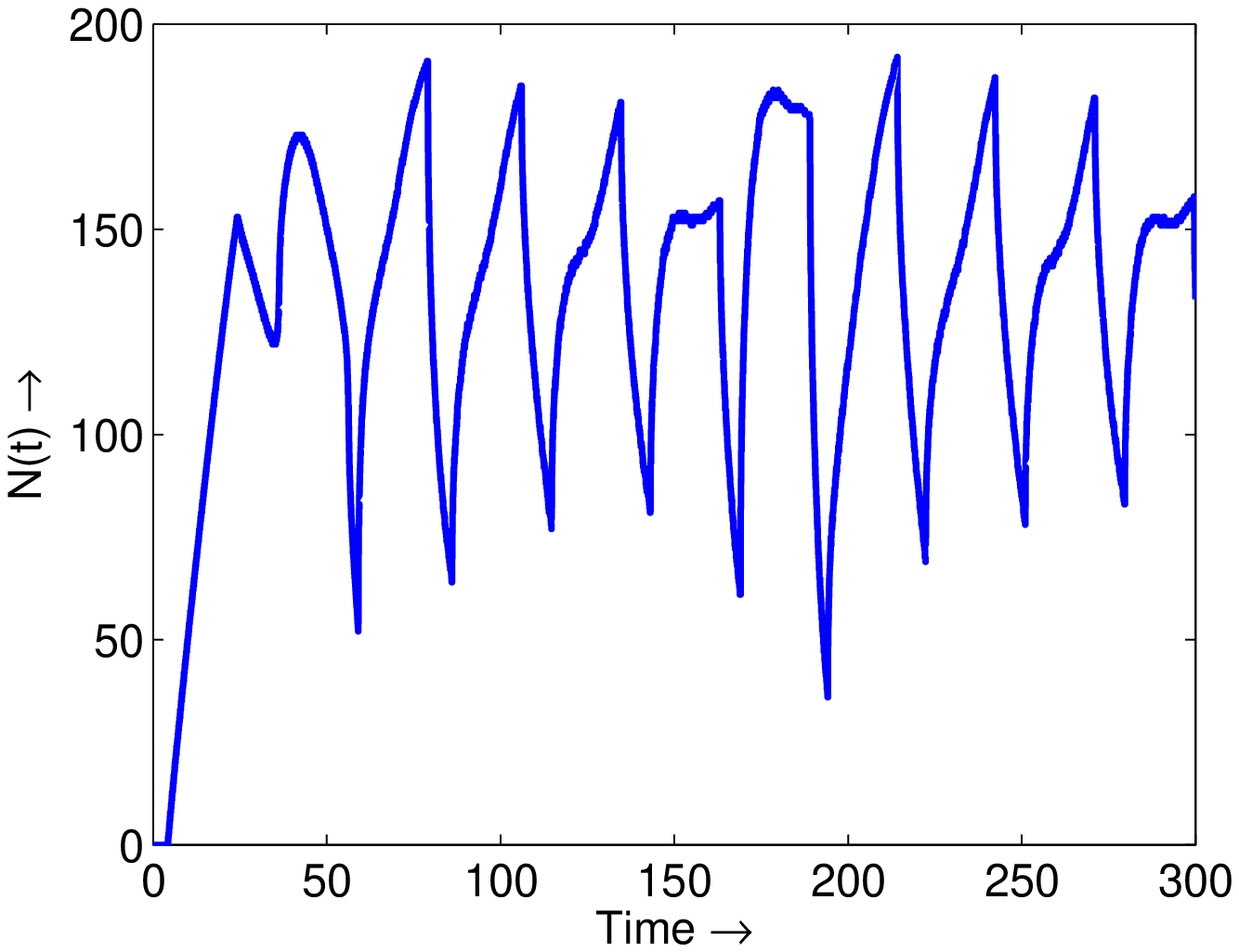}
 \hspace*{-0.5cm}
\includegraphics[scale=0.37]{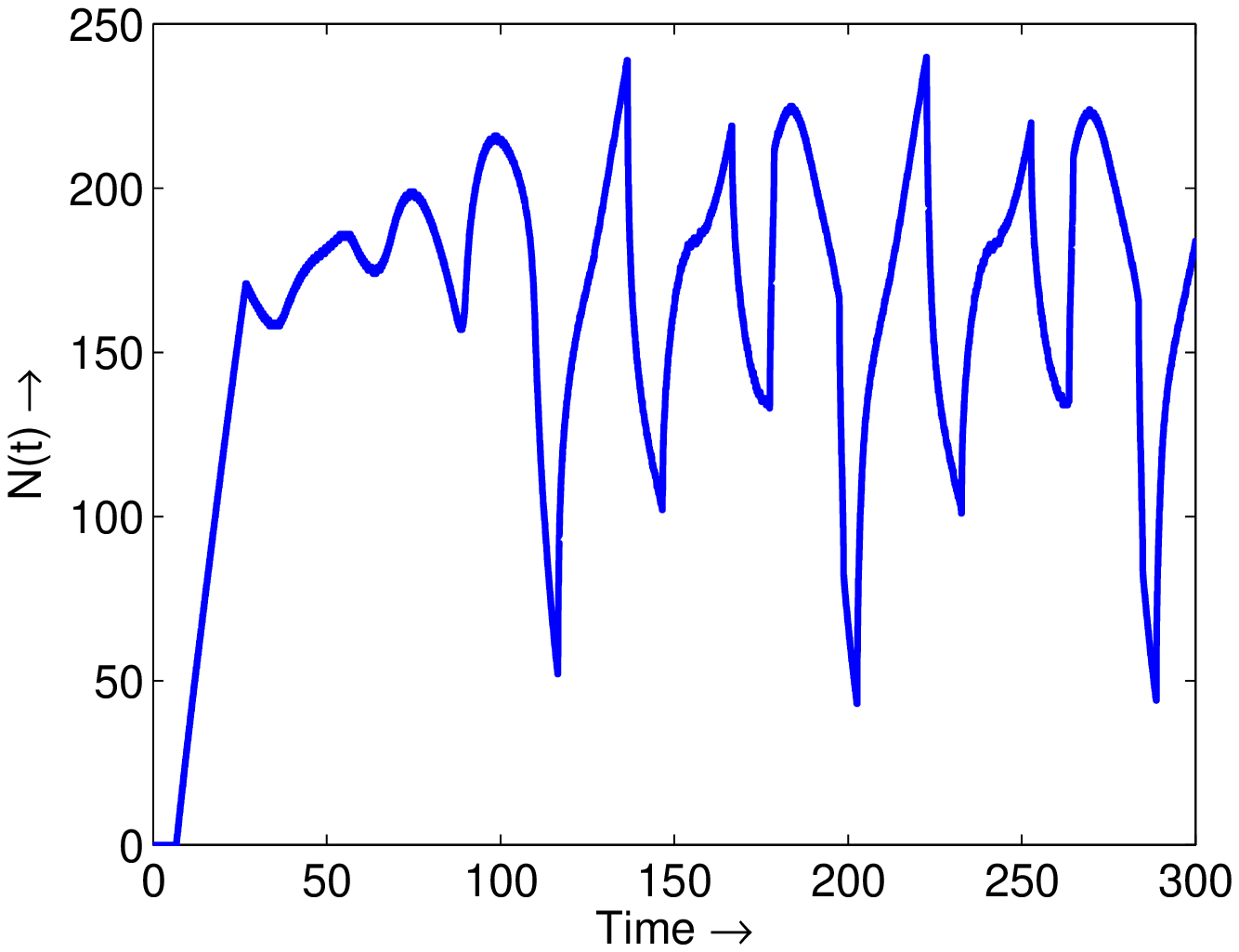}}
\caption{The number of differentiated cells $N(t)$ for a constant initial condition $p_1^0=p_2^0=0.1$.
$N_0=150, \alpha=0.003$ (left), $N_0=150, \alpha=0.009$ (middle), $N_0=200, \alpha=0.003$ (right).}
\label{det1}
\end{figure}

Figure \ref{det1} shows numerical simulations in the case of a fixed (deterministic) initial condition.
The number of differentiated cells oscillates around the target value $N_0$. Therefore the separatrix oscillates
and the initial condition $(p_1^0,p_2^0)$ belongs to different basins of attraction during different time intervals.
These oscillations do not result from the instability of a stationary solution since the stationary solution does not exist.

%%%%%%%%%%%%%%%%%%%%%

%\newpage

\subsection{Random initial conditions}

\subsubsection{Existence of stationary solutions}

We indicated in the previous section that the problem with fixed deterministic initial conditions does not have
stationary solutions. We will consider now random initial conditions.

We consider system (\ref{mod1}) with the functions $F_1$ and $F_2$
given by equalities (\ref{mod2}), (\ref{mod3}). The initial
conditions are taken in domain $D$ defined in (\ref{mod4}) with a
uniform distribution. For each given $N$, denote by $R_N(t)$ the
number of differentiated cells produced by the system during the
time interval $[t-T,t]$. Since the initial conditions are random,
this number can also have random oscillations. Let $\widehat R_N$ be
its average value with respect to asymptotically large time
interval,

$$ \widehat R_N = \lim_{S \to \infty} \frac1S \int_0^S R_N(t) dt . $$
We will consider that the problem has a stationary solution if the following equality holds:

\begin{equation}\label{mod5}
  \widehat R_N = N.
\end{equation}
We will show that this equation has an approximate solution.

Consider system  (\ref{mod1}) for this value of $N$. Separatrix $S$ of the point
$P_3$, that is the trajectory which approach this saddle point, separates basins of attraction of the points $P_1$ and $P_2$.
Denote by $D_i(N)$ the part of the domain $D$ which belongs to the basin of attraction of the point $P_i$, $i=1,2$.
Let $|D_i(N)|$ be measures of these subdomains. Then $\gamma_N = |D_2(N)|/|D|$ is the proportion of differentiated cells for asymptotically
large time, and $\gamma_N T/\tau$ is their average number during time interval $T$. Equation

\begin{equation}\label{mod6}
  \gamma_N T/\tau = N
\end{equation}
can be considered as approximation of equation (\ref{mod5}). We will indicate conditions when this equation has a solution.
Let us note first of all that $\gamma_N$ is a decreasing function of $N$. The value of $N$ changes from $0$ to $T/\tau$.
If system (\ref{mod1}) is well defined for all these values of $N$ and possesses the same structure (number and stability of stationary points),
then equation (\ref{mod6}) has a unique solution.

%\medskip
%\noindent
%{\color{blue} It can be specified in what sense $\gamma_N T/\tau$ is the average number of differentiated cells.}
%
%

%%%%%%%%%%%%%%%%%

\subsubsection{Stable stationary solutions}

\begin{figure}[htbp]
\centerline{\includegraphics[scale=0.22]{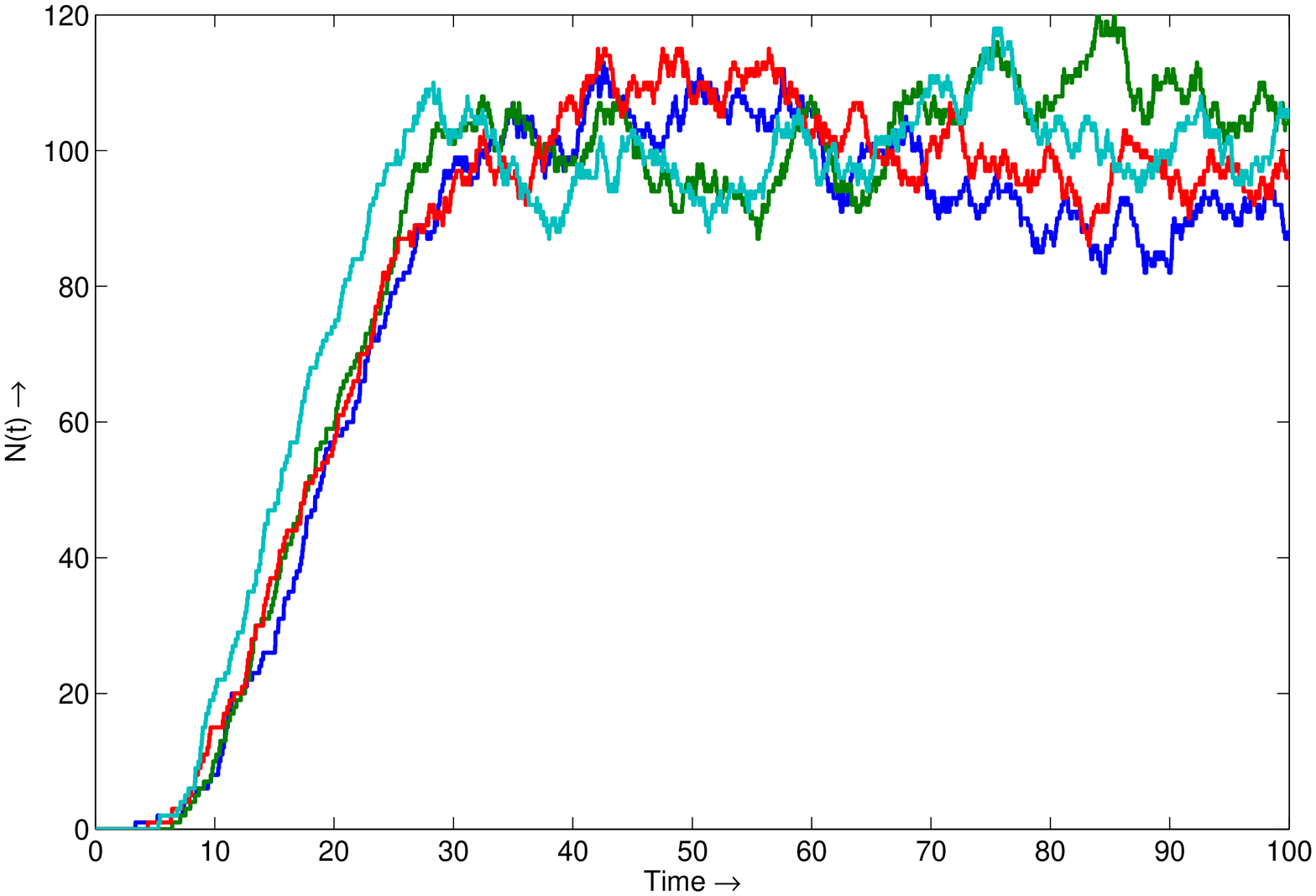}
 \hspace*{-0.5cm}
\includegraphics[scale=0.22]{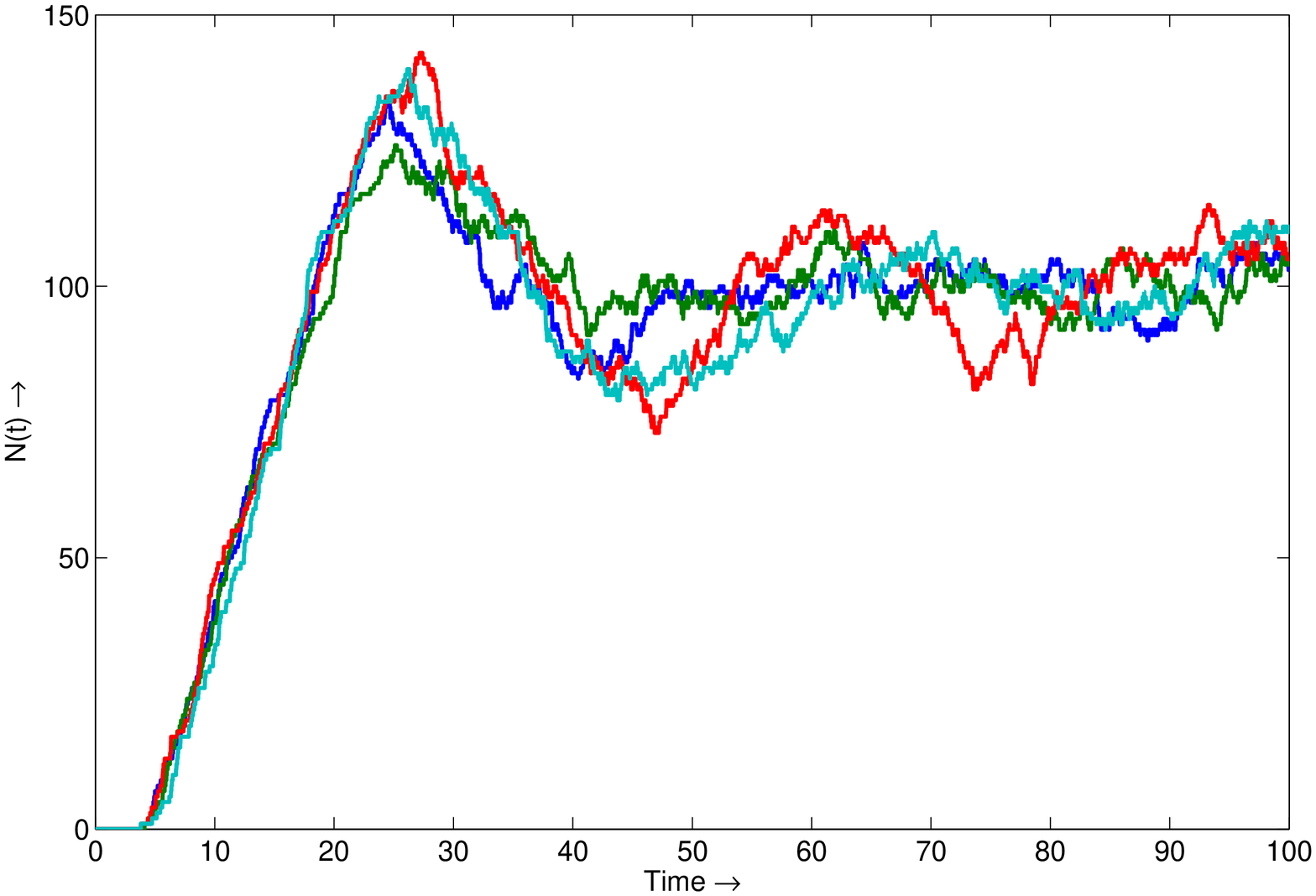}
 \hspace*{-0.5cm}
\includegraphics[scale=0.22]{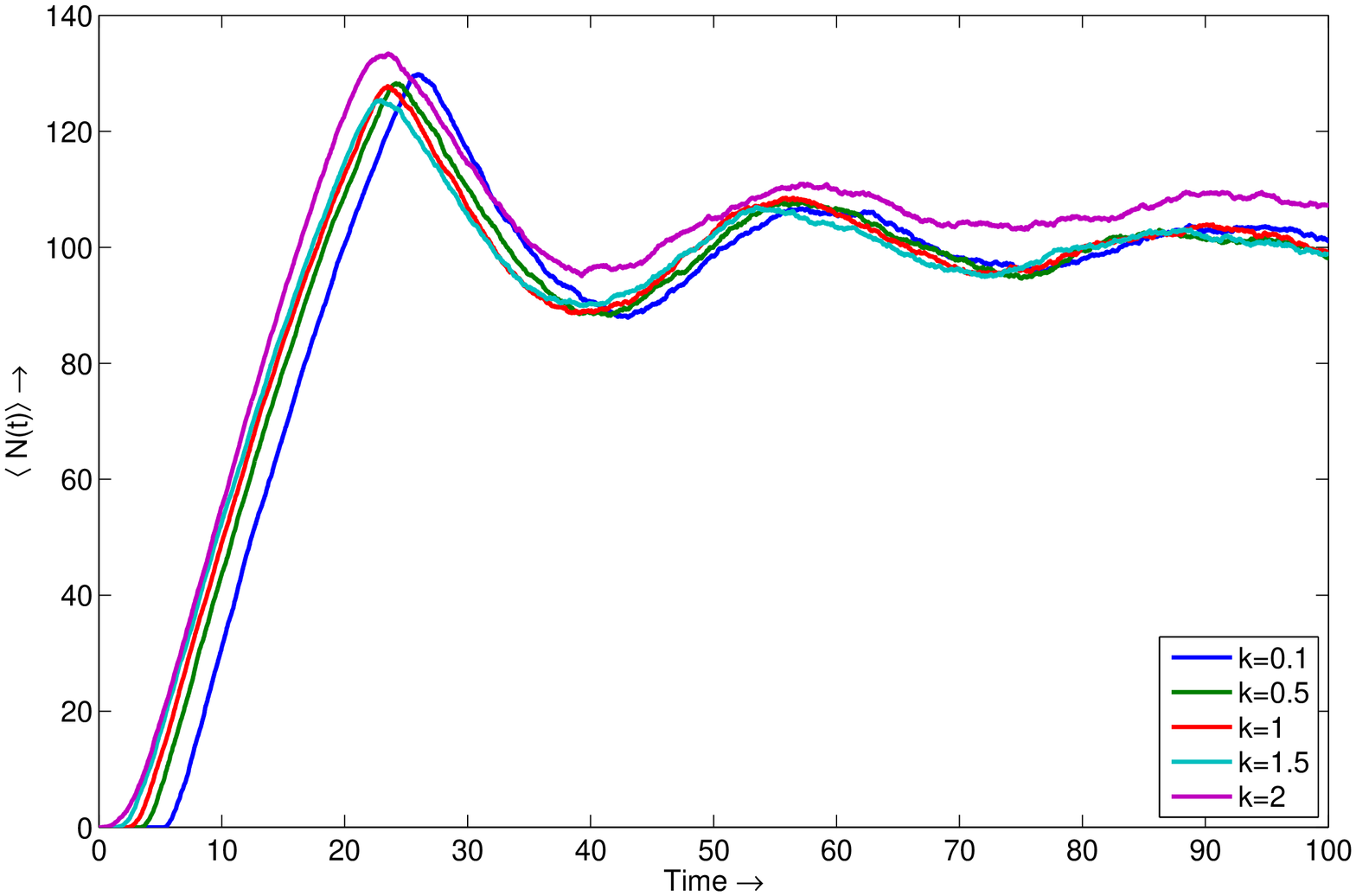}}
\caption{The number of cells $N(t)$ as a function of time for $N_0=100$, $\alpha=0$ (left) and $\alpha=0.009$ (middle).
Four curves show different simulations with the same values of parameters. The curves in the right figure
correspond to different values of $k$. Average value of $N(t)$ for 50 simulations,
$N_0=100$, $\alpha=0.009$.}
\label{fig2}
\end{figure}

In this section the initial conditions $p_1^0$ and $p_2^0$ will be taken from the square domain
$0 \leq p_1^0, p_2^0 \leq k$ as a random variable with a uniform distribution.
During the time interval $T$ there are $T/\tau$ new cells $A$. In the case of symmetric parameters and
initial conditions, a half of them will differentiate. For the values of parameters considered here, we obtain $N=100$ differentiated cells $B$.
Figure \ref{fig2} shows examples of numerical simulations for $N_0=100$ and different values of $\alpha$. If $\alpha$ is sufficiently large,
then there is an overshoot in cell number. Its value for large time converges to $N_0$ after small decaying oscillations.

\begin{figure}[ht!]
\centerline{\includegraphics[scale=0.22]{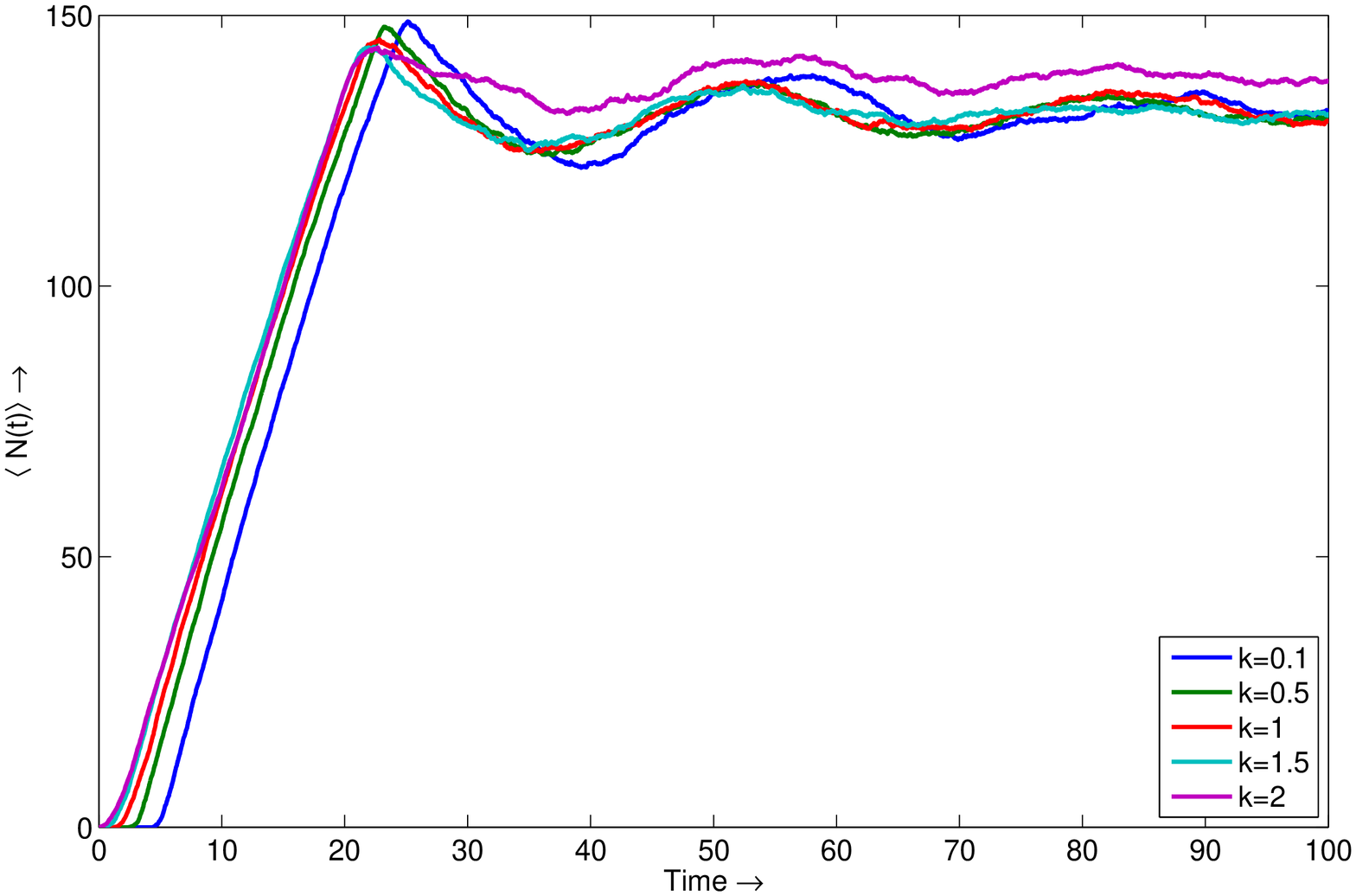}
 \hspace*{-0.5cm}
\includegraphics[scale=0.22]{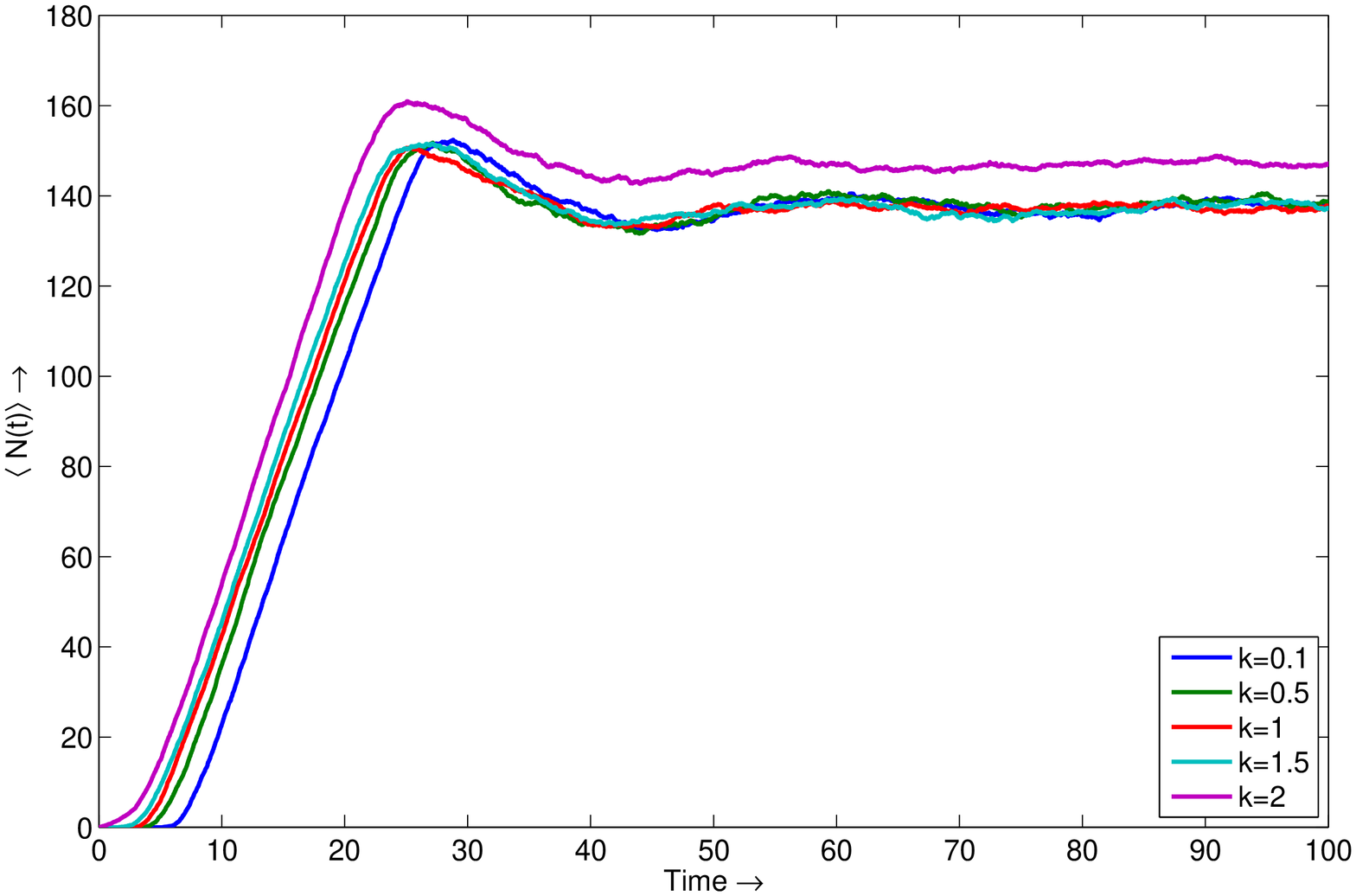}
 \hspace*{-0.5cm}
\includegraphics[scale=0.22]{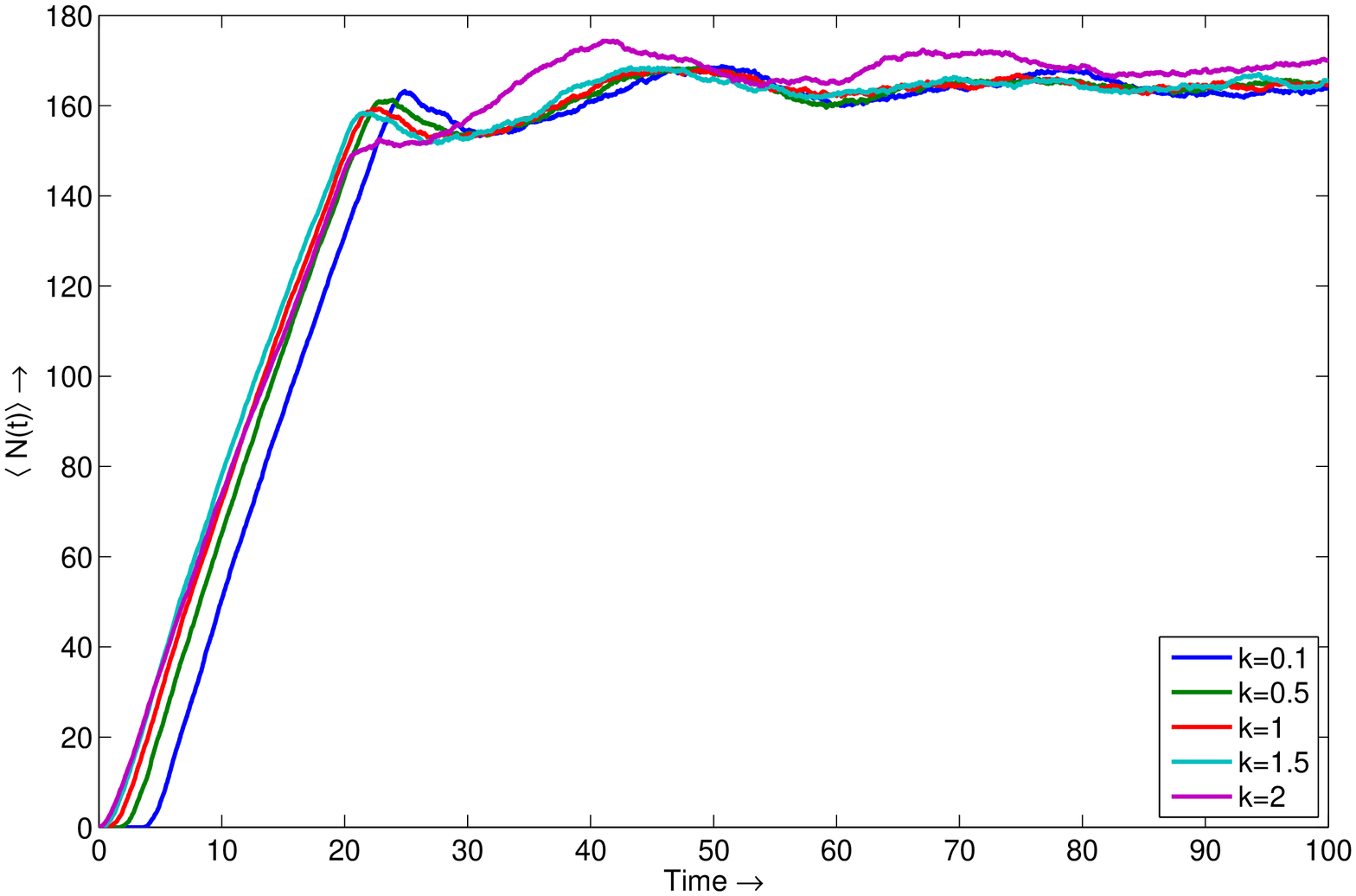}}
\caption{The number of differentiated cells $N(t)$ as a function of time for $N_0=150$, $\alpha=0.009$ (left),
$N_0=200$, $\alpha=0.003$ (middle) and $\alpha=0.009$ (right) and different values of $k$. Average for 50 simulations.}
\label{sym3}
\end{figure}

Figures \ref{fig2} and \ref{sym3} show the evolution of the  number of differentiated cells $N(t)$ in time for different
values of parameters. In all cases it converges to some limiting value $N_\infty$.
It is interesting to note that the function $N(t)$ is practically independent
of the value of $k$ which determines the choice of initial conditions. In the wide range of variation of this parameter,
$0.1 \leq k \leq 1.5$, the curves coincide. A slight difference appears for $k=2$.

%\begin{figure}[ht!]
%\centerline{\includegraphics[scale=0.3]{./figures/alphadot009_N0100.eps}
% \hspace*{-0.5cm}
%\includegraphics[scale=0.3]{./figures/alphadot009_N0150.eps}}
%\caption{The number of differentiated cells $N(t)$ as a function of time for
%$N_0=100$ (left) and $N_0=150$ (right), $\alpha=0.009$ and different values of $k$. Average for 50 simulations.}
%\label{sym29}
%\end{figure}

\begin{figure}[ht!]
\centerline{\includegraphics[scale=0.45]{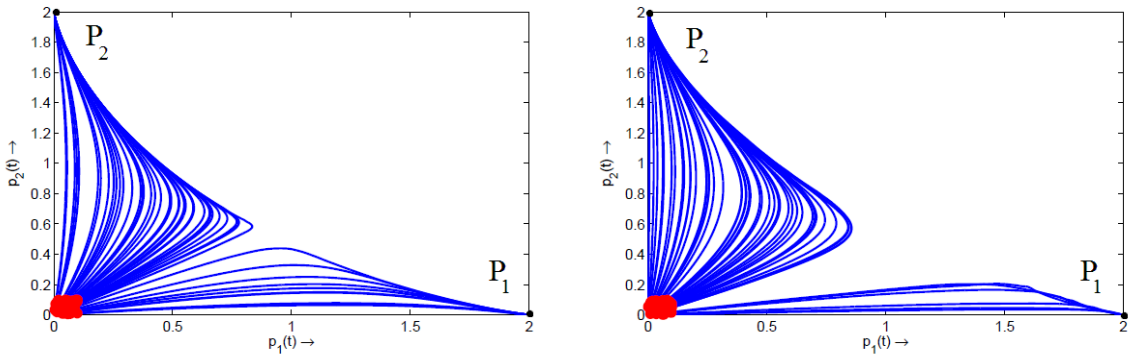}}
%\centerline{\includegraphics[scale=0.35]{trajectories_alpha_dot003_IC_dot1.eps}
% \hspace*{-0.5cm}
%\includegraphics[scale=0.35]{trajectories_alpha_dot009_IC_dot1.eps}}
\caption{Trajectories of system (\ref{mod1}) for the same simulations as in Figure \ref{sym3} and
for $95 < t < 100$. The domain of initial conditions (red) is split by the separatrix of the saddle point $P_3$
in two subdomains. Depending on the initial condition, trajectories approach either point $P_1$ or $P_2$;
$N_0=150$ (left), $N_0=200$ (right).}
\label{sym31}
\end{figure}

Let us recall that in the case of deterministic initial conditions (Section 3.1), $N(t)$ does not converge to a stationary value.
Hence we can expect appearance of oscillations for very small values of $k$. We will return to this question in the next section.
Therefore convergence to a stationary solution is determined by randomness in the initial conditions.
 We illstrate this mechanism in Figure \ref{sym31}. The domain $D$ of initial conditions is split by the separatrix
into two subdomains which belong to the basins of attraction of the points $P_1$ and $P_2$. Depending on the position of the initial condition,
the corresponding trajectory will converge to one of these stationary points. The areas of these two subdomains determine the proportion
of differentiated cells and the value $N_\infty$. Let us note that system of equations (\ref{mod1})-(\ref{mod3}) has time dependent
coefficients. However for $t$ sufficiently large, $N(t)$ is close to a constant, and the curves in Figure \ref{sym31} can be
considered as trajectories of the autonomous system.

The global regulation by the number of differentiated cells $N(t)$ through the coefficients of the system
(see (\ref{mod3})) influences the limiting value $N_\infty$.
It increases with the increase of $N_0$ (Figures \ref{sym3}, \ref{sym31}) but in general it is different from $N_0$.
Clearly, if $N(t) \equiv N_0$, then from (\ref{mod3}) we have $a_{21} = a_{21}^0$. In this case, the system is symmetric and
exactly half of the total cell number differentiate. For the values of parameters under consideration $N_\infty=100$.
Therefore, if $N_0 \neq 100$, then $N_\infty < N_0$. The value of $N_\infty$ increases and approaches $N_0$ when we take
greater values of $\alpha$ (Figure \ref{sym3}).

The simulations presented above are carried out for a fixed value  $a_{21}^0$ in (\ref{mod3}).
We can also introduce its dependence on  $N$. Then we can obtain a better approximation of
$N_0$ by $N_\infty$. We varied $a_{21}^0$ in large limits and obtained convergence to a stationary value $N_\infty$ which
depends on the value of $a_{21}^0$ (not shown here).

Thus global regulation allows us to obtain a given stationary value of the number of cells for a large interval of variation
of the coefficients and of the range of random initial conditions.

\begin{figure}[htbp]
\centerline{\includegraphics[scale=0.37]{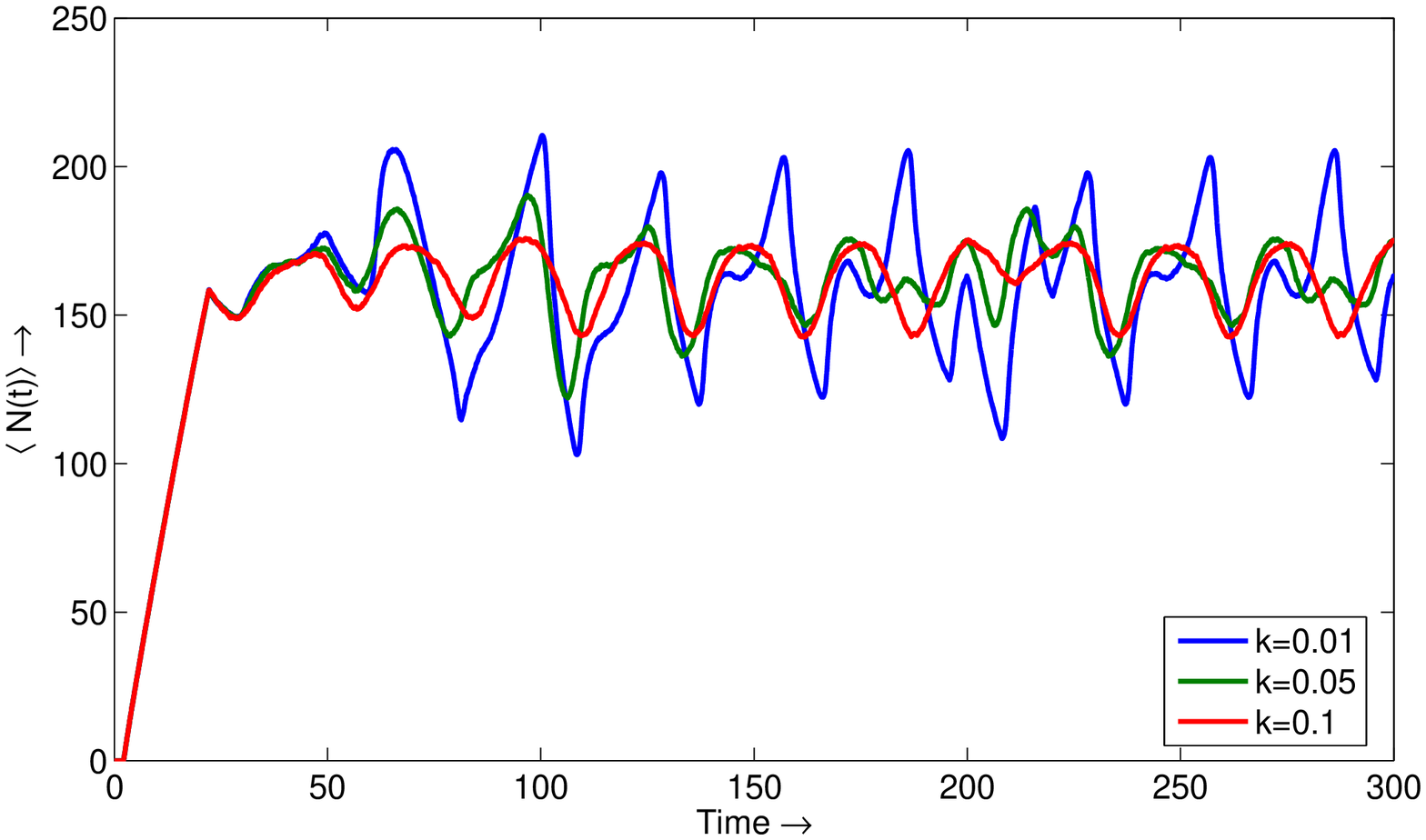}
\hspace*{-0.5cm}
\includegraphics[scale=0.37]{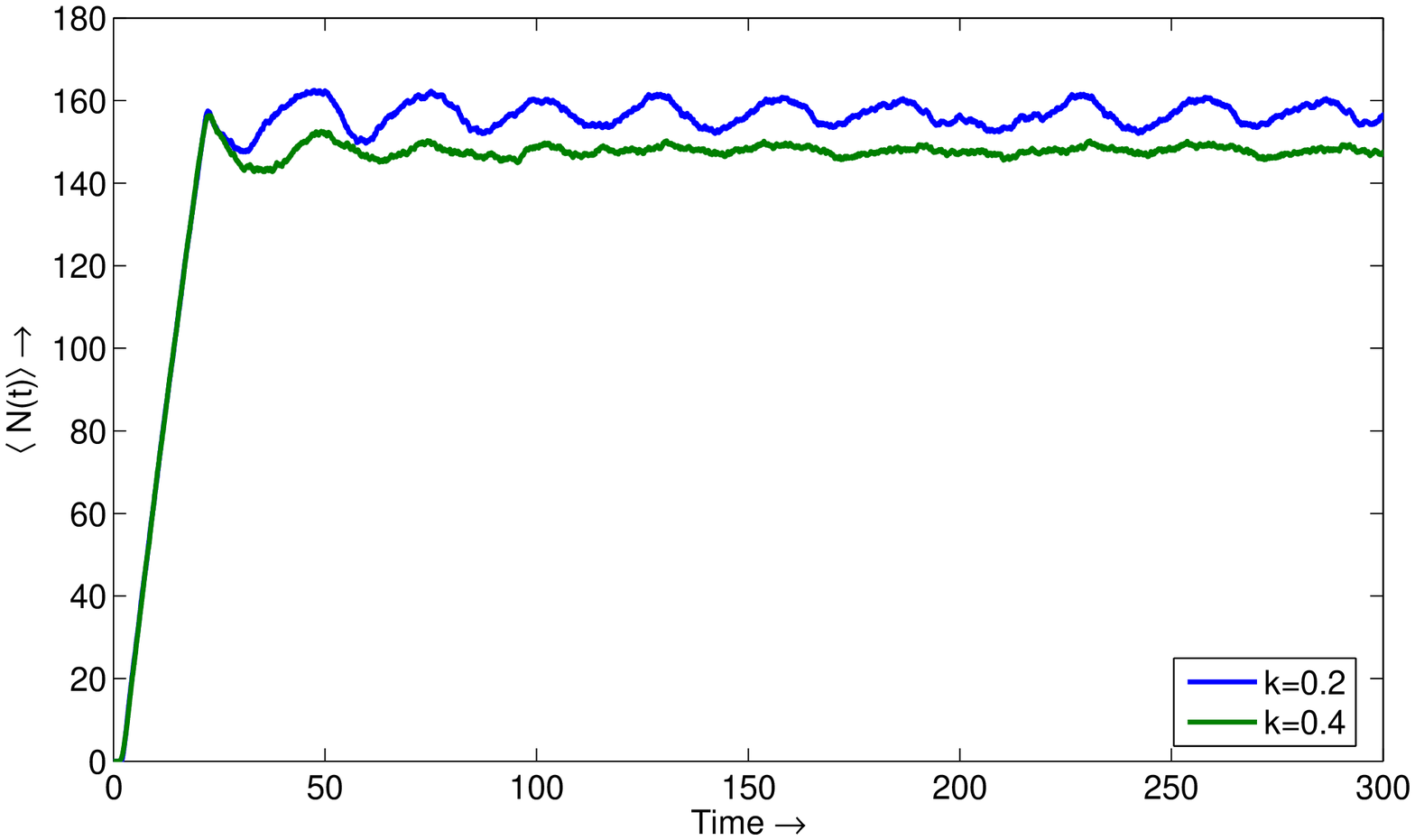}}
\caption{The number of differentiated cells $N(t)$ for
$N_0=150$, $\alpha=0.009$, $a=0.4, b=0.6$ and different values of $k$. When $k$ increases,
the amplitude of oscillations decay.}
\label{fig-as10}
\end{figure}

%%%%%%%%%%%%%%%%

\subsubsection{Oscillating solutions}

In Section 3.2.1 we considered deterministic initial conditions and observed periodic oscillations of the number of
differentiated cells. In Section 3.2.2 we studied the case with random initial conditions from a sufficiently large domain.
In this case, the number of differentiated cells converges to a constant value.

Let us now consider transition between these two cases.
Initial conditions will be uniformly distributed in the domain $[a-k,a+k] \times [b-k,b+k]$.
Figure \ref{fig-as10} shows the number of differentiated cells in time for different values of $k$ and all
other parameters fixed. The amplitude of oscillations decreases when $k$ increases, and they practically disappear
for $k=0.4$. Hence randomness of initial conditions can suppress oscillations. We will discuss a possible mechanism of
this stabilization in Section 4.

\begin{figure}[htbp]
\centerline{\includegraphics[scale=0.37]{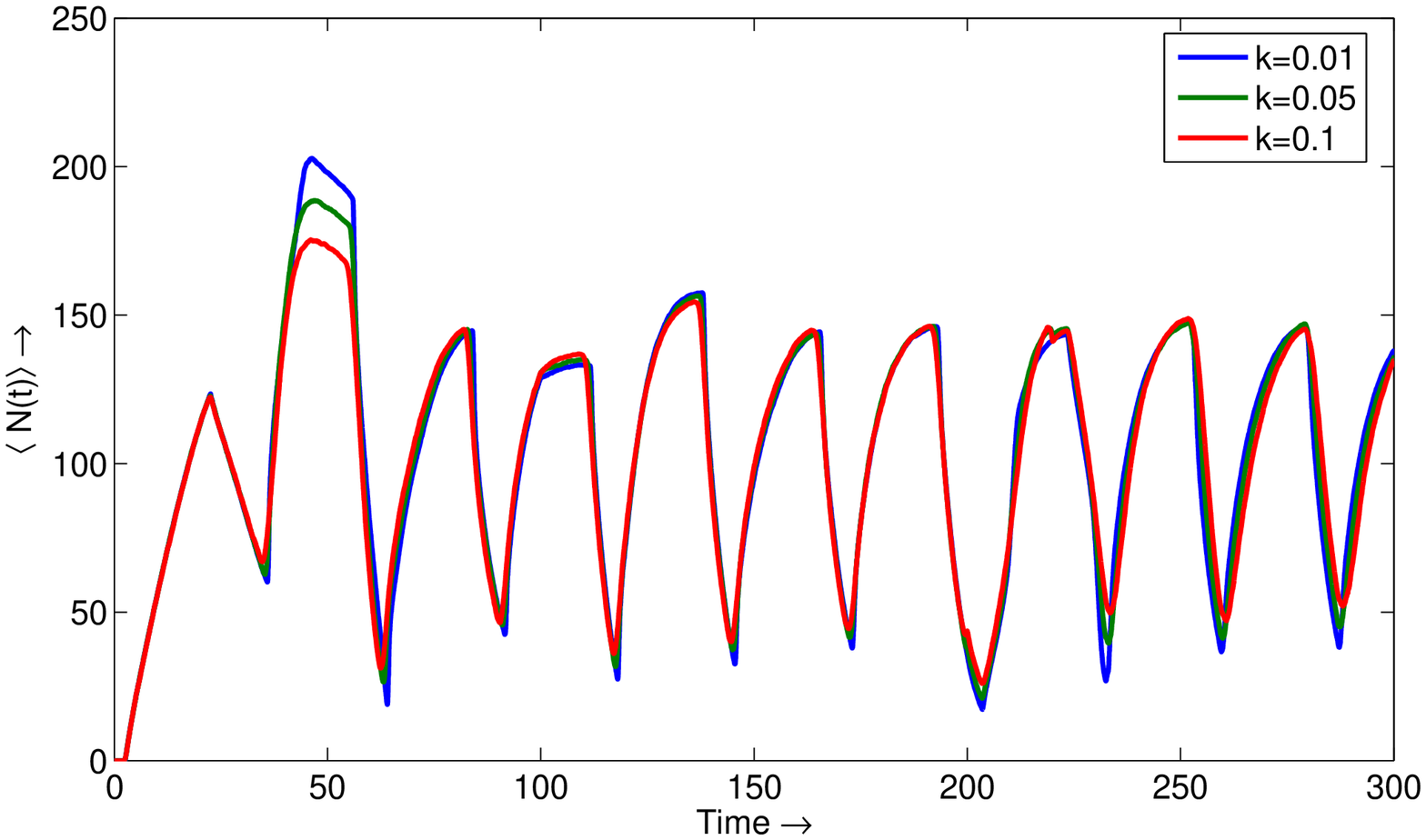}
\hspace*{-0.5cm}
\includegraphics[scale=0.37]{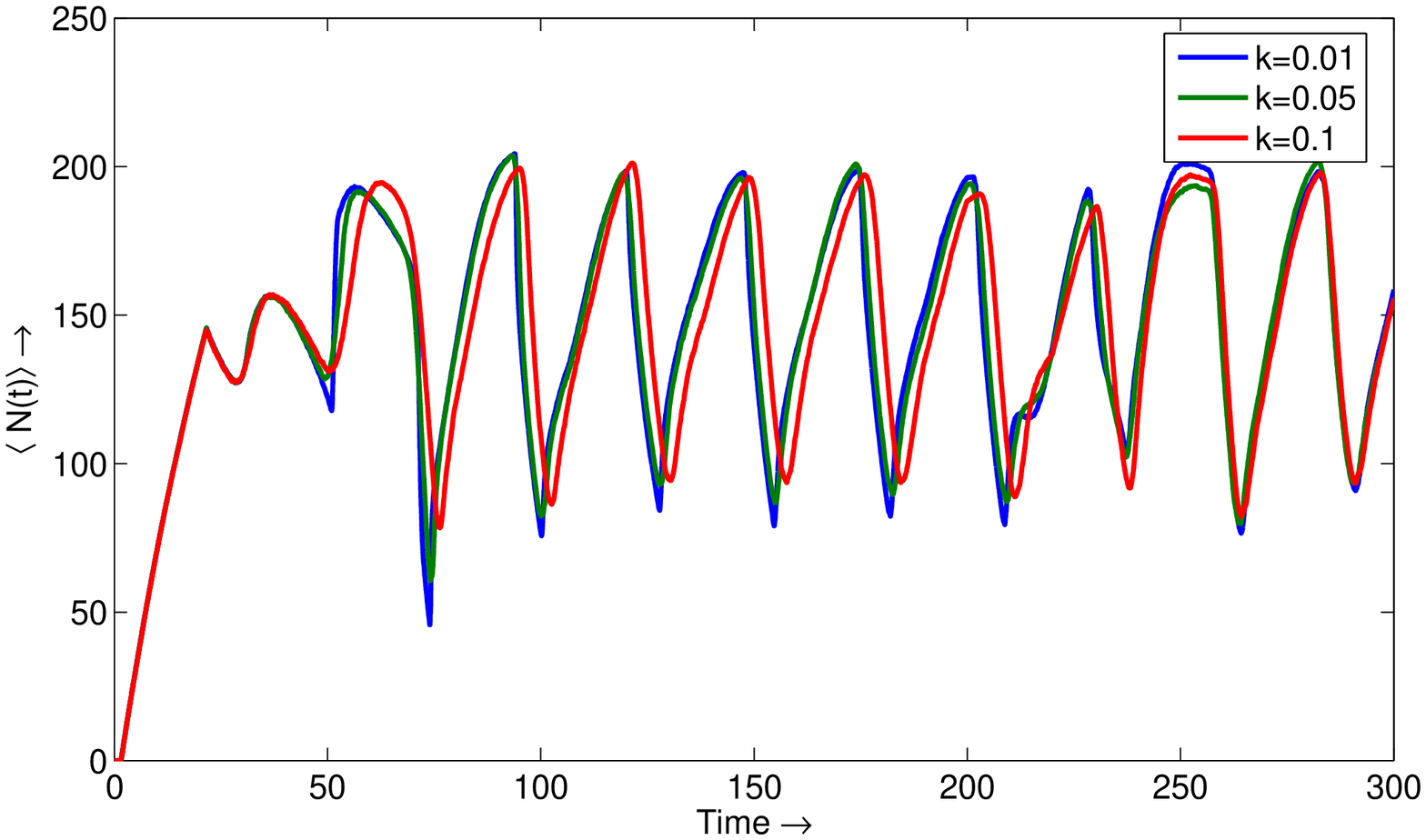}}
\caption{The number of differentiated cells $N(t)$ for
$N_0=150$ (left), $N_0=200$ (right), $\alpha=0.009$, $a=1, b=0.2$. The oscillations are independent of
the value of $k$ in the considered range.}
\label{fig-as1}
\end{figure}

\begin{figure}[ht!]
\centerline{\includegraphics[scale=0.45]{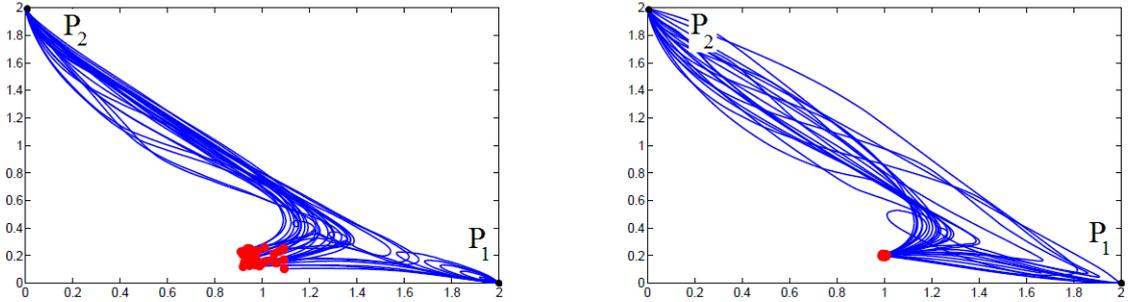}}
%\centerline{\includegraphics[scale=0.3]{trajectories_AssymIC_a1bdot2_kdot1.eps}
%\includegraphics[scale=0.3]{trajectories_AssymIC_a1bdot2_kdot01.eps}}
%\centerline{\includegraphics[scale=0.35]{trajectories_AssymIC_a1bdot2_kdot01.eps}}
\caption{Curves $p_1(t),p_2(t)$ on the $(p_1,p_2)$-plane for the same values of parameters as in the previous figure.
Initial conditions are shown as red dots.}
\label{fig-as2}
\end{figure}

If we now take other values of $a$ and $b$ (Figure \ref{fig-as1}), then the oscillations are practically
independent of the value of $k$ (in the considered range). Persistence of the oscillations can be related to large time delay
for cells to differentiate. Indeed, the time which the trajectory needs to reach the critical value $p_2^*$ depends on the
initial condition.

System (\ref{mod1}) has time dependent coefficients (\ref{mod3}).
Therefore the extracellular regulation through $N(t)$ acts on the cell during its all life time.
Indeed, some of the curves $(p_1(t),p_2(t))$ shown in Figure \ref{fig-as2} cross each other because the system
is not autonomous. This extracellular regulation
can have more influence on the intracellular concentrations than the choice of initial conditions.
This is why the results shown in Figure \ref{fig-as1}
are practically independent of the range where the initial conditions are chosen from.

\begin{figure}[ht!]
\centerline{\includegraphics[scale=0.8]{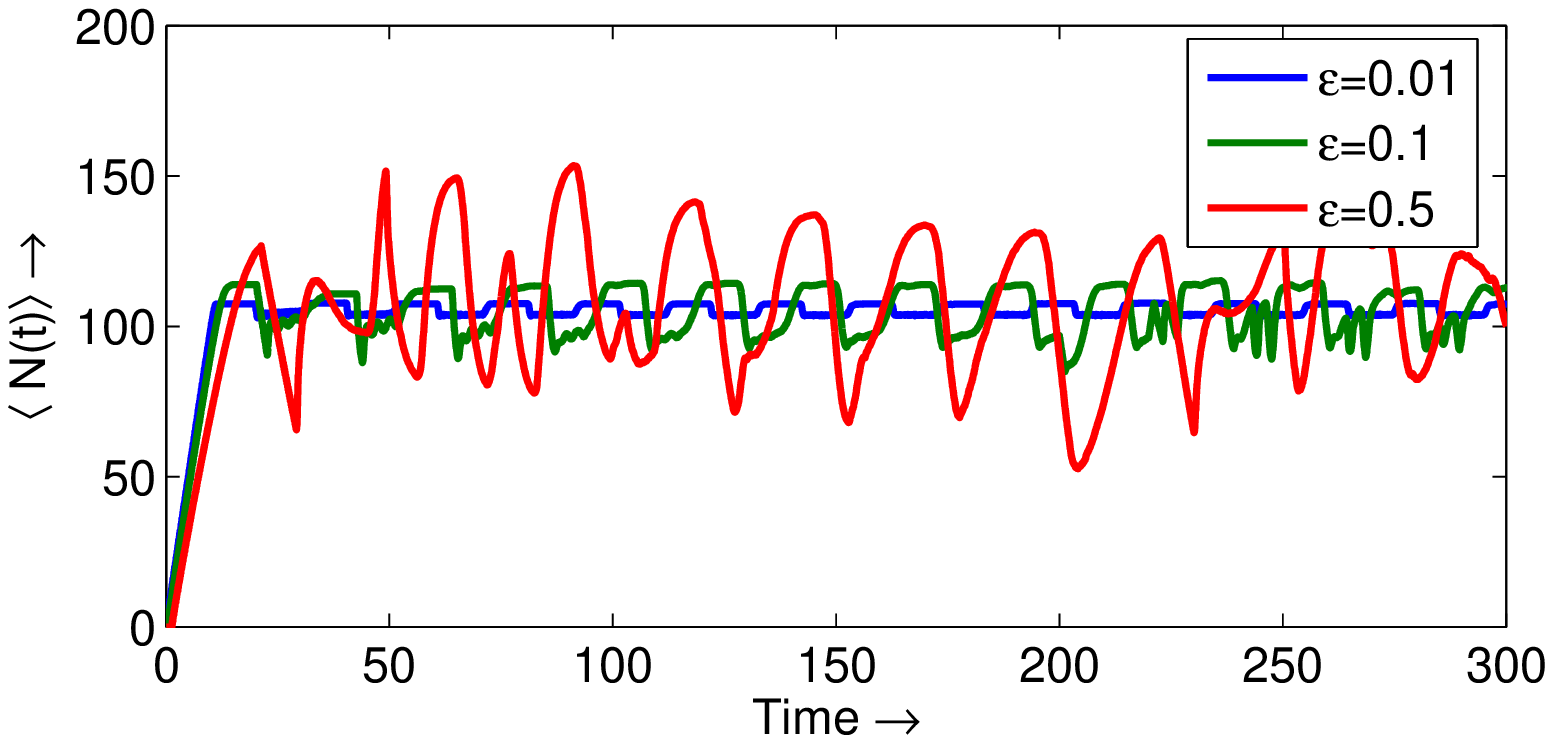}}
\caption{The number of differentiated cells $N(t)$ for
$N_0=150, \alpha=0.009$, $k=0.01$ and different values of $\epsilon$ in system (\ref{e1}), (\ref{e2}).}
\label{fig-as7}
\end{figure}

In order to analyze how time oscillations are related to the time delay of the intracellular regulation,
we introduce a small parameter $\epsilon$ in system (\ref{mod1}):

\begin{equation}
\epsilon \label{e1} \frac{dp_1}{dt} = k_1 p_1 (1 - a_{11} p_1 - a_{12} p_2),
\end{equation}

\begin{equation}\label{e2}
\epsilon \frac{dp_2}{dt} = k_2 p_2 (1 - a_{21} p_1 - a_{22} p_2).
\end{equation}
Decreasing $\epsilon$ we accelerate intracellular regulation. The amplitude of oscillations also decreases
(Figure \ref{fig-as7}). This confirms the hypothesis that time oscillations are related to time delay as it is the case
in delay differential equations (Section 4).

%%%%%%%%%%%%%%%%%%%%%%%%%%%%%%%%%%%%%%%

%\newpage

\setcounter{equation}{0}

\section{Analytical approximation}

The total number $N(t)$ of differentiated cells can be found from the equation

\begin{equation}\label{aa1}
  \frac{dN}{dt} = H(t) - M(t) ,
\end{equation}
where $H(t)$ and $M(t)$ the rates of cell birth and death at time $t$.
%This equation is similar to the second equation
%in system (\ref{sys1}) written for the cell number instead of cell concentrations.
Since cells have a fixed life time $T$, then

$$ M(t) = H(t-T) , $$
and equation (\ref{aa1}) can be written as follows:

\begin{equation}\label{aa2}
  \frac{dN}{dt} = H(t) - H(t-T) .
\end{equation}
In the numerical model considered above, instead of the birth rate $H(t)$ we consider cell choice between differentiation and apoptosis of cells $A$.
Therefore $H(t)$ is the rate of appearance of newly differentiated cells at time $t$.
It is determined by solutions of the system of differential equations for the intracellular variables. So we cannot express
it through $N$ in order to have a closed equation for this variable. However we can derive such equation if we do some simplifying assumptions.

Let us recall that the choice between differentiation and apoptosis is determined by the intracellular variables $p_1$ and $p_2$ described
by system (\ref{mod1}). Its solution depends on the initial conditions and on the values of the coefficients which depend on $N(t)$
due to the extracellular regulation.

 Let us fix some $t=t_0$. The value $N(t_0)$ determines the values of the coefficients of system (\ref{mod1}). Basins of attraction
of stable points $P_1$ and $P_2$ are separated by the stable manifold of the unstable point $P_3$. Denote the subdomains of $D$ these basins of attractions by
$D_1$ and $D_2$, respectively. If the initial condition is located in $D_1$, the cell will go in the direction of the point $P_1$ and will
die by apoptosis, if it is in $D_2$, then it will go in the direction of $P_2$ and will differentiate. If we know the ratio of initial
conditions in $D_1$ and $D_2$, then we can try to estimate the rate of differentiation. However, it cannot be done directly for the model under
consideration because:
a) The coefficients of the system depend on time through $N(t)$. Hence a trajectory, which is initially in the basin of attraction of one
of the two stable points, can change it and go to another point,
b) Even inside the same basin of attraction and for fixed values of the coefficients, the time to reach the critical values $p_1^*$ and
$p_2^*$ depends on the initial condition. This time increases when the initial condition approaches the stable manifold of the point $P_3$.
Hence the cell appeared at time $t=t_0$ can differentiate or die at any other time $t > t_0$.
c) The initial condition is chosen randomly from some given domain $D$.

Therefore we cannot reduce the model with intracellular regulation described by ordinary differential equations to a closed delay
equation (\ref{aa2}).
Consider now a simplified model where a cell appeared at time $t_0$ makes its choice between differentiation and apoptosis exactly
at the moment $t=t_0$ of its appearance. This choice is not influenced by further change of $N(t)$. Then for each cell we consider system
(\ref{mod1})  with constant coefficients. These coefficients can be different for cells appeared at different moments of time.
In this case, the probability for the initial condition to be in domains of attraction of each of two stable stationary points at time $t_0$ are given by the formula

$$ \pi_1(t_0) = |D_1|/|D| , \;\;\; \pi_2(t_0) = |D_2|/|D| , $$
where $|D|$ is the area of the domain $D$, $|D_i|$ is the area of the corresponding subdomain. Since the coefficients of system
(\ref{mod1}) depend on time, then the subdomains $D_i$ are also time dependent, $D_i = D_i(t_0)$.

Next, let us assume that the time to reach the critical values $p_1^*$ and $p_2^*$, when the cell differentiates or dies, does not depend on the
initial condition. This means that instead of the ordinary differential system of equations we consider a binary (or Boolean) mode with some
given time delay $\tau_0$. If the initial condition belongs to $D_1(t_0)$, then the cell will differentiate at time $t=t_0+\tau_0$,
otherwise it will die. Then the rates of differentiation and apoptosis are given by the relations

$$ d(t) = \gamma \pi_1(t-\tau_0) , \;\;\; a(t) = \gamma \pi_2(t-\tau_0) , $$
where $\gamma$ is the rate of appearance of new cells. It remains to note that $D_i(t)$ are some given functions of $N(t)$. Hence

$$ d(t) = \gamma f(N(t-\tau_0)) , \;\;\; a(t) = \gamma (1 - f(N(t-\tau_0)) . $$
Let us note that $f(N)$ is a decreasing function: if there are more differentiated cells, then the rate of their production is less.

Thus, equation (\ref{aa2}) can be written in the form

\begin{equation}\label{aa3}
 \frac1\gamma \;  \frac{dN}{dt} = f(N(t-\tau)) - f(N(t-T-\tau))  .
\end{equation}
This equation is obtained as an approximation of equation (\ref{aa2}) under some simplifying assumptions.
It can also be considered independently, as a simple phenomenological model of cell population under
extracellular regulation. The first term in the right-hand side of this equation determines the influx of cells in the
tissue and the second term their death.
Equation (\ref{aa3}) is equivalent to the integral equation

$$ N(t) = \gamma \int_{t-T}^t f(N(s-\tau)) ds + n_0 , $$
where $n_0$ is an arbitrary constant. We note that $n_0=0$ since $N(t)$ equals exactly the number of cells differentiated from
$t-T$ to $t$ given by the integral in the right-hand side.
The last equation can also be written in the form:
$$ N(t) = \gamma \int_0^T f(N( t - y -\tau)) dy  .  $$

If we look for a constant solution $N(t)=N$, then we obtain the equation

$$ N = \gamma T f(N) . $$
Since $f(N)$ is  a positive decreasing function, then it has a unique solution $N=N^*$
(cf. Section 3.2.1).

%%%%%%%%%%%%%%%%%%%%%%%

\paragraph{Linear stability analysis.}

Let us linearize equation (\ref{aa3}) about the constant solution $N^*$:

\begin{equation}\label{aa4}
 \frac1\gamma \;  \frac{du}{dt} = f'(N^*) ( u(t-\tau)) - u(t-T-\tau))  .
\end{equation}
We look for its solution as $u(t) = \exp(\lambda t)$. Then we get

\begin{equation}\label{aa41}
  \frac1h \; \lambda = e^{-\lambda \tau} - e^{-\lambda (T+ \tau)} ,
\end{equation}
where $h = \gamma f'(N^*)$.
We note that $\lambda=0$ is a solution of this equation. It is related to the fact that equation (\ref{aa3}) has a family of solutions.
We will look for the solution $\lambda=i \phi$. Then we get the system of two equations

$$ \cos(\phi \tau) = \cos(\phi(T+\tau)) , $$

$$ - \frac1h \phi = - \sin(\phi \tau) + \sin(\phi(T+\tau)) . $$
From this system of equations we can find $\phi$ and the stability boundary.
From the first equation we get

$$ \phi(T+\tau) = 2 \pi n - \phi \tau  $$
(sinus of these two values should have opposite signs). Then from the second equation we have

\begin{equation}\label{aa5}
  \sin x = \frac{x}{2 h \tau} ,
\end{equation}
where

$$  x = \frac{2 \pi n \tau}{T + 2 \tau} \; . $$
Equation (\ref{aa5}) determines the stability boundary. It has a nonzero solution if $\tau > 1/(2h)$.

%%%%%%%%%%%%%%%%%%%%%%%
%
%\paragraph{Existence of periodic solutions ?}

\paragraph{Implications for the original problem.}

In the analytical model we obtain that oscillations appear if the
time delay $\tau$ of the intracellular regulation is sufficiently
large. This conclusion is in agreement with the result of the
numerical simulations presented in Section 3. Indeed, let the
initial conditions be uniformly distributed in the square
$[a-k,a+k]\times[b-k,b+k]$. If the point $(a,b)$ is more close to
the point $P_2$, the time delay to reach the critical value $p_1^*$
is greater than in the case where it is close to $P_1$. Accordingly,
if $a=1, b=0.2$ the oscillations are strong (Figure \ref{fig-as1}),
for $a=0.4, b=0.6$ they are weaker (Figure \ref{fig-as10}), for
$a=0.1, b=1$ no oscillations (not shown).

We can also use this approximate analytical result to understand the influence of stochasticity.
When $k$ is small and $a=0.6, b=0.6$, there are oscillations. When the initial conditions are taken
from the square $[0,k]\times[0,k]$, $k=1,1.5$, there are no oscillations. When we take a big square, then there are
different initial conditions. For some of them time delay is small (as in the case where $(a,b)$ is close to $P_1$).
Such solutions with small time delay can suppress oscillations (see the next paragraph).

%We can also analyze the influence of other parameters, $T$, $\alpha$, $\epsilon$.

%%%%%%%%%%%%%%%%%%%%%%%

\paragraph{Two delays in the intracellular regulation.}

In order to understand the role of variation of initial conditions
in the intracellular regulation, we can consider two production
functions $f$ and $g$ with different delays:

\begin{equation}\label{aa6}
 \frac1\gamma \;  \frac{dN}{dt} = f(N(t-\tau_1)) - f(N(t-T-\tau_1))  + g(N(t-\tau_2)) - g(N(t-T-\tau_2))  .
\end{equation}
This equation corresponds to the case of two different initial conditions with two different time delays.
We can expect that if the first delay is large enough to produce oscillations considered along, and the second delay is short,
then it can remove the oscillations. Let us verify this conjecture. After linearization we obtain the equation for the eigenvalues:

\begin{equation}\label{aa7}
  \lambda = h_1 \left( e^{-\lambda \tau_1} - e^{-\lambda (T+ \tau_1)} \right)  + h_2 \left(
e^{-\lambda \tau_2} - e^{-\lambda (T+ \tau_2)} \right) ,
\end{equation}
where $h_1 = \gamma f'(N^*)$, $h_2 = \gamma g'(N^*)$, $N^*$ is a solution of the equation

$$ N = \gamma T [f(N) + g(N)] . $$
We set $\tau_1=\tau$ as in equation (\ref{aa41}) and
substitute in the right-hand side of equality (\ref{aa7}) the solution $\lambda=i\phi$ of equation (\ref{aa41}). Then we get

\begin{equation}\label{aa8}
  \lambda = \frac{h_1}{h} \; i \phi +  h_2 \;
  \left( e^{- i \phi \; \tau_2} - e^{-i \phi \; (T+ \tau_2)} \right) \; .
\end{equation}
Denote by $R$ the real part of the expression in the right-hand side of this equality. Then

 $$ R = h_2(\cos (\phi \tau_2) - \cos (\phi (T + \tau_2))) . $$
If $\tau_2=0$, then $R \leq 0$ since $h_2 < 0$. Moreover, $R=0$ only if $\phi T = 2 \pi n$ for some integer $n$.
If we do not take into consideration these exceptional cases, then $R < 0$ for $\tau_2=0$ and, by continuity,
$R$ remains negative for sufficiently small $\tau_2$.

Thus, if we take the same values of parameters which determine the
stability boundary in the case of one time delay and introduce the
second (short) time delay, then we get in the stability region.
Hence the second time delay stabilizes the stationary solution. This
model with two delays confirms the results of the numerical
simulations which show that random initial conditions can suppress
oscillations. Among random initial conditions, those which
correspond to large time delay, lead to the oscillations, while the
initial conditions, which correspond to small time delay, can
suppress the oscillations.

\section{Discussion}

The goal of this work is to study how intracellular regulation, extracellular
regulation and stochasticity in the initial intracellular concentrations determine cell fate.
The results of this work allow us to suggest that they act in the following way.
Intracellular regulation  provides a choice between two or several options.
Global extracellular regulation controls the realization of this choice.
Stochasticity in the initial conditions stabilizes the system.

Each of these three elements should satisfy certain conditions. Intracellular regulation should possess
some kind of bistability. If we describe it with a system of ordinary differential equations,
this can be two stable stationary points which correspond to different cell fates. However bistabilty
can be understood here in a larger sense. The phase space should be split into two invariant manifolds
by a central manifold. The invariant manifolds correspond to different cell fates. The central manifold
should be controlled by the global extracellular regulation through the coefficients of the
ordinary differential system of equations in order to adapt the volume of invariant manifolds to get
the target value of cell number.

Intracellular regulation is extremely complex and in many cases there is only a partial information about it.
An important methodological question is how to take it into account
if only a part of this network is known and the reaction constants are not known. From the point of view
of the modelling presented in this work, we should take the known part of the intracellular regulation and complete
it to the system which described the required cell fate. This additional part, since it is not known, can be
constructed in different ways. However, this difference may not be essential from the point of view of the cell
population because the global extracellular regulation will control it. This control will be
different for different intracellular regulations but it will finally produce the same result.

%We can make an analogy with a traveller who should go from a point A to a point B. There are different pathway
%between these points, and he may not know them in advance. However, knowing the final destination and using the distance control
%to the point $B$, he will arrive there. And this is not very important which pathway he took.

%A related question is why to describe the details of the intracellular regulation if we can use a simplest bistable system.
%If we want to act on the intracellular regulation in order to change the dynamics of the cell population, we can introduce
%the known part of the pathway and act on it. In the previous example with the traveller, we can block the pathways
%to the point B, and travellers will go to other destinations. This is how many drugs act. We can evaluate their action even
%without knowing the complete regulatory network.

We will finish this discussion with the last element of our regulatory system, stochasticity in the initial conditions.
We showed in numerical simulations and in analytical models that it can stabilize the system. Oscillations is an intrinsic
property of cellular systems because of time delay in their regulation. So stochasticity can be one of possible mechanisms
to suppress these oscillations. In this case an interesting question is about the origin of stochasticity. It can result from random
perturbations, from stochastic dynamics in the case of small number of molecules or
from some underlying mechanism which can produce some variations in initial protein concentrations.
The question about possible mechanisms producing variation of initial condition is completely open.

\end{document}